\documentclass[aps,prl,twocolumn,superscriptaddress]{revtex4}%
\usepackage{amssymb}
\usepackage{amsfonts}
\usepackage{amsmath}
\usepackage{graphicx}%
\providecommand{\U}[1]{\protect\rule{.1in}{.1in}}

\begin{document}
\title{Superconductivity at the vacancy disorder boundary in K$_{x}$Fe$_{2-y}$%
Se$_{2}$ }
\author{Chunruo Duan}
\affiliation{University of Virginia, Charlottesville, VA 22904, USA. }
\author{Junjie Yang }
\affiliation{University of Virginia, Charlottesville, VA 22904, USA. }
\author{Yang Ren}
\affiliation{Advanced Photon Source, Argonne National Laboratory, Argonne, Illinois 60439,
USA }
\author{Despina Louca$^{\ast}$}
\affiliation{University of Virginia, Charlottesville, VA 22904, USA. }

\begin{abstract}
The role of phase separation in the emergence of superconductivity in alkali metal doped iron selenides A$_{x}$Fe$_{2-y}$Se$_{2}$ (A = K, Rb, Cs) is revisited. High energy X-ray diffraction and Monte Carlo simulation were used to investigate the crystal structure of quenched superconducting (SC) and as-grown non-superconducting (NSC) K$_{x}$Fe$_{2-y}$Se$_{2}$ single crystals. The coexistence of superlattice structures with the in-plane $\sqrt{2}\times\sqrt{2}$ K-vacancy ordering and the $\sqrt{5}\times\sqrt{5}$ Fe-vacancy ordering were observed in SC and NSC crystals along side the \textit{I4/mmm} Fe-vacancy free phase. Moreover, in the SC crystal an Fe-vacancy disordered phase is additionally present. It appears at the boundary between the \textit{I4/mmm} vacancy free phase and the \textit{I4/m} vacancy ordered phase
($\sqrt{5}\times\sqrt{5}$). The vacancy disordered phase is most likely the host of superconductivity.

\end{abstract}
\date{\today}

\maketitle

The self-organization of electronic nematic states and multi-phase separation are at the heart of the underlying lattice complexity prevalent in the high-temperature iron-based and cuprate superconductors \cite{kasahara}. Superconductivity emerges by suppressing the static antiferromagnetic (AFM) order \cite{dai} but spin and charge density fluctuations are commonly inferred and may be responsible for electron pairing. Coupled with these fluctuations is a heterogeneous lattice where the spatial interplay between spin and charge yields nanoscale phase separation \cite{bianconi2}. Therefore, the nature of the lattice structure is key to elucidating the symmetry-breaking ground state that may lead to superconductivity. The A$_{x}$Fe$_{2-y}$Se$_{2}$ system is a test bed for exploring the very peculiar states that appear with the close proximity of superconductivity to a magnetic insulating state \cite{louca}, leading to a multiphase complex lattice.

The A$_{x}$Fe$_{2-y}$Se$_{2}$ (A = K, Rb, Cs) iron selenide superconductor has been intensely studied \cite{guo} in part due to the possible role of Fe-vacancy order and whether or not phase separation occurs between SC and NSC regions \cite{wbao,wbao2,yan,ding,scott}. With vacancies at both the A and Fe sites, a well-known structural transition occurs when the Fe vacancies order at $T_{S}\sim580\ K$\cite{wbao}. In the high temperature tetragonal phase with the \textit{I4/mmm} space group, the vacancies are randomly distributed at both the Fe and A sites. Upon cooling below $T_{S}$, a superlattice structure appears due to Fe vacancy order. Several scenarios have been proposed regarding the nature of the crystal structure below $T_{S}$. In one, the lattice is phase separated into a minority \textit{I4/mmm} phase which is compressed in-plane and extended out-of-plane in comparison to the high temperature centrosymmetric phase and has no Fe vacancies, and a majority \textit{I4/m} phase with the Fe vacancies ordered in different superlattice patterns \cite{ricci,ricci2,li,yuan}. The most commonly reported superlattice structure with Fe vacancy order is the $\sqrt{5}\times\sqrt{5}\times1$ \cite{wbao,wbao2,ye,zavalij}. More recently, other superlattice patterns have been reported in the literature such as the $2\times2\times1$\cite{wbao2,wang}, the $1\times2\times1$\cite{wbao2,yan,wang,kazakov} and the $\sqrt{8}\times\sqrt{10}\times1$\cite{ding}.

The distinction among the different superlattice patterns arises from the underlying order of the Fe and alkali metal sublattices. In the \textit{I4/m} phase, the Fe site symmetry is broken from the high temperature \textit{I4/mmm} space group, giving rise to two crystallographic sites. Preferred site occupancy leads to the $\sqrt{5}\times\sqrt{5}$ supercell, in which one site is empty (or sparsely occupied) while the other is almost full. Magnetic ordering is characteristic of this phase. Below $T_{N}\sim560\ K$, AFM order arises in the \textit{I4/m} phase that persists well below $T_{c}$\cite{wbao}. The AFM state, commonly reported in the literature\cite{wbao,yan,ye}, is robust unlike what has been observed in all other Fe-based superconductors, and its coexistence with the superconducting state has raised concerns about the validity of the s+/- coupling mechanism coupled with the absence of hole pockets at the Fermi surface and the lack of nesting in this system.\cite{zhang} More recently, evidence of alkali site vacancy order has been presented as well with a $\sqrt{2}\times\sqrt{2}$ superlattice structure within the \textit{I4/mmm} phase in K$_{x}$Fe$_{2-y}$Se$_{2}$\cite{ricci} and Cs$_{x}$Fe$_{2-y}$Se$_{2}$\cite{bosak,taddei,porter}. The centrosymmetry of the \textit{I4/mmm} is broken due to the alkali metal order. The \textit{I4/mmm} phase with no Fe vacancy has largely been attributed to be the host of superconductivity in part because of the absence of magnetism and vacancies at least at the Fe site. 

It is understood at present that by post-annealing and quenching, superconductivity can be enhanced in this system \cite{han,Ryu} even though the actual mechanism remains unknown. Magnetic refinement from neutron powder diffraction measurements revealed that magnetic order does not exclude the presence of a SC phase\cite{junjie}. Moreover, the SC shielding fraction obtained from bulk magnetic susceptibility measurements correlates to a larger volume fraction of the \textit{I4/m} phase instead of the smaller volume fraction of the \textit{I4/mmm} phase. To investigate this issue further, high energy X-ray scattering measurements were performed on two kinds of K$_{x}$Fe$_{2-y}$Se$_{2}$ single crystals, one annealed and SC, and the other as-grown and NSC. In combination with Monte Carlo simulation, it is shown that superconductivity in the quenched crystal is most likely present in regions between the $\sqrt{5}\times\sqrt{5}\times1$ \textit{I4/m} domain boundaries, bordering the \textit{I4/mmm} domains with no Fe vacancies. Thus superconductivity in this system appears at the crossover of the vacancy order-disorder transition. Quenching increases the boundary walls around the \textit{I4/m} domains, leading to an increase of the percolation paths and an enhancement of superconductivity.

\begin{figure}[ptb]
	\includegraphics[width=0.5\textwidth]{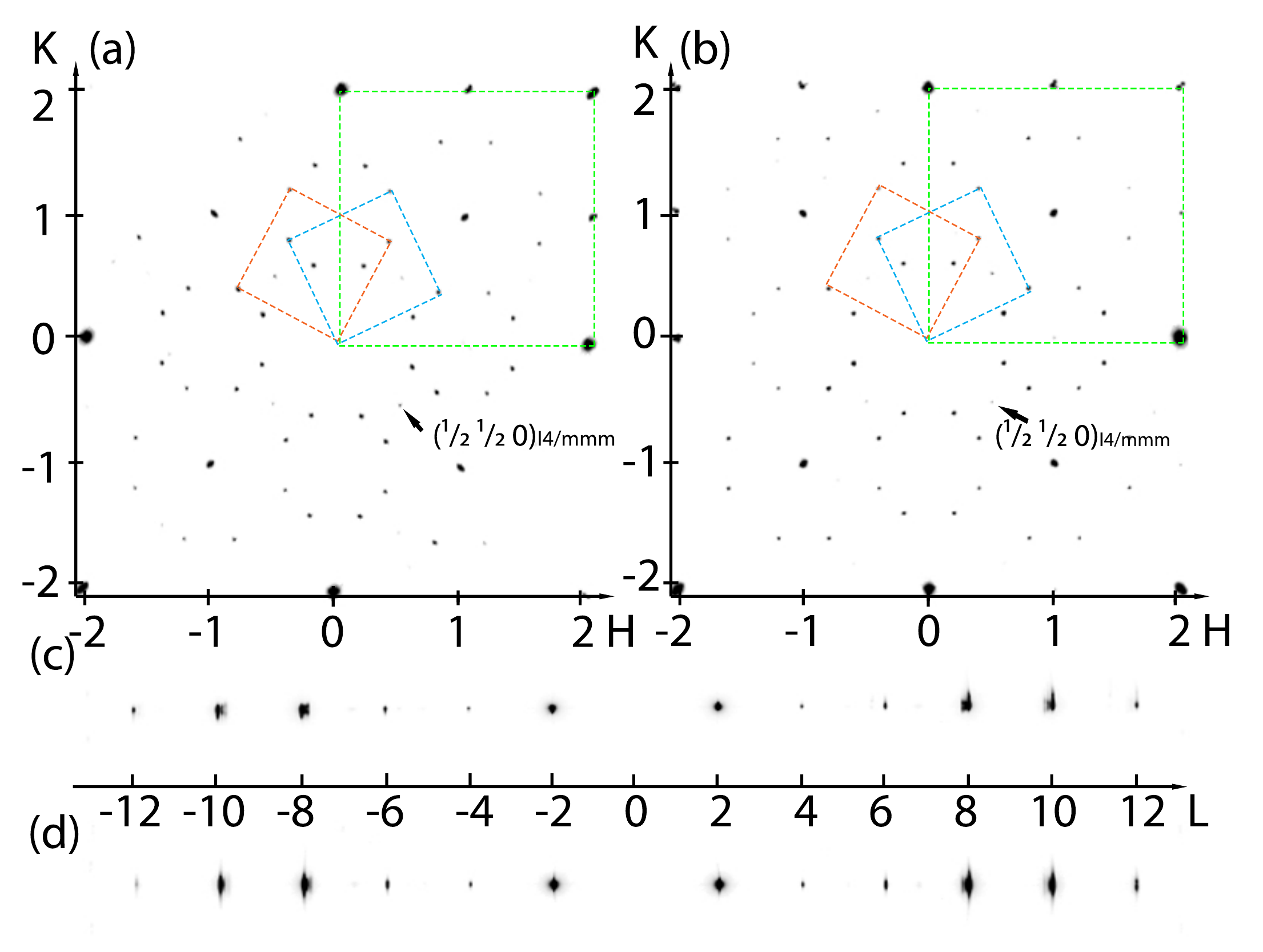} \caption{The diffraction patterns from the \textit{hk0} plane from (a) as-grown and (b) quenched crystals. In (c) and (d) are teh plots along the 00l direction in the as-grown and quenched crystal, respectively. The \textit{hk0} patterns consist of two configurations of the \textit{I4/m} phase highlighted by the two inner dashed boxes, and the \textit{I4/mmm} phase with a $\sqrt{2}\times\sqrt{2}$ superlattice structure highlighted by the outer dashed box.}%
	\label{pkcmp}%
\end{figure}

\begin{figure}[ptb]
	\includegraphics[width=0.5\textwidth]{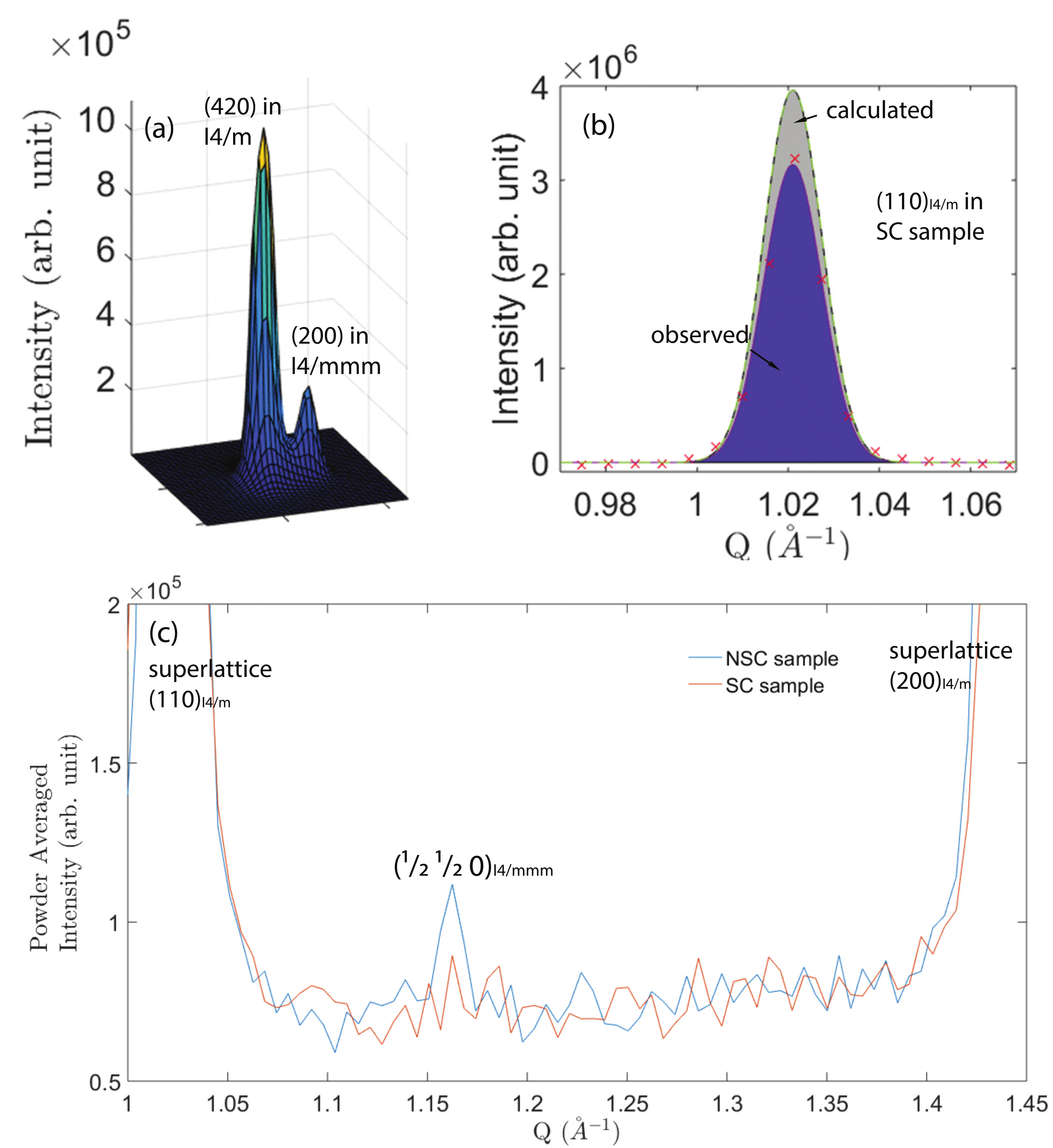} \caption{(a) A comparison between Bragg peaks from \textit{I4/m} and \textit{I4/mmm} phases. (b) The observed superlattice peak (110) in SC crystal is compared with the calculated intensity based on an ideal \textit{I4/m} structure. (c) The powder integral of the \textit{hk0} scattering plane in the vicinity of the superlattice peak ($\frac{1}{2}$ $\frac{1}{2}$ 0).}%
	\label{fig2}%
\end{figure}

Single crystals of K$_{x}$Fe$_{2-y}$Se$_{2}$ were grown using the self-flux method. The first step of the synthesis involved the preparation of high-purity FeSe by solid state reaction. Stoichiometric quantities of iron pieces (Alfa Aesar; 99.99\%) and selenium powder (Alfa Aesar; 99.999\%) were sealed in an evacuated quartz tube, and heated to 1075 $^{\circ}$C for 30 hours, then annealed at 400 $^{\circ}$C for 50 hours, and finally quenched in liquid nitrogen. In the second step, a potassium grain and FeSe powder with a nominal composition of K:FeSe = 0.8:2 were placed in an alumina crucible and double-sealed in a quartz tube backfilled with ultrahigh-purity argon gas. All samples were heated at 1030 $^{\circ}$C for 2 hours, cooled down to 750 $^{\circ}$C at a rate of 6 $^{\circ}$/hr, and then cooled to room temperature by switching off the furnace. High quality single crystals were mechanically cleaved from the solid chunks. In the final step, the annealed crystals were additionally thermally treated at 350 $^{\circ}$C under argon gas for 2 hours, followed by quenching in liquid nitrogen. The crystals that were not heat treated were labeled as-grown. The magnetic susceptibility and transport were measured from 2 to 300 K and the as-grown crystal is NSC while the annealed crystal is SC. Back-scattered scanning electron microscopy (SEM) measurements were carried out at room temperature on the two samples\cite{junjie}. The characterization of these crystals was previously reported in Ref.\cite{junjie}. The SEM measurements showed that the surface morphology of the as-grown crystal has two kinds of regions: rectangular islands with a bright color and a background with a dark color. On the other hand, instead of island-like domains, very small bright dots were observed on the surface of the annealed crystal. The single crystal diffraction measurements were carried out at the Advanced Photon Source of Argonne National Laboratory, at the 11-ID-C beam line. In-plane and out-of-plane measurements were carried out on both types of crystals at room temperature.

The X-ray diffraction from the \textit{hk0} scattering plane shows evidence of coexistence of multiple phases. Shown in Figs. 1(a) and 1(b) are the patterns corresponding to the as-grown and quenched crystals, respectively. Several features are observed in both that arise from the presence of the two configurations of $\sqrt{5}\times\sqrt{5}\times1$ superlattice structure with the \textit{I4/m} symmetry \cite{ricci,zavalij} indicated by the two inner dashed boxes as well as the \textit{I4/mmm} phase indicated by the outer dashed box. Indicated by an arrow is a superlattice peak indexed as ($\frac{1}{2}$ $\frac{1}{2}$ 0). The lattice constant calculated from the peak position matches that of the \textit{I4/mmm} phase, indicating a $\sqrt{2}\times\sqrt{2}$ A-site vacancy ordered structure in the \textit{I4/mmm} phase. The scattering patterns along the \textit{l}-direction are shown in Figs. 1(c) and (d) for the as-grown and quenched crystals, respectively. Bragg peaks from \textit{I4/mmm} appear at the lower $Q$ side of the \textit{I4/m} peaks. No l=2n+1 superlattice peaks are observed along the (00l) direction, leaving the out-of-plane stacking of the $\sqrt{2}\times\sqrt{2}$ K-vacancy order unclear. Due to sample rotation during measurement, weak reflections are observed at the lower $Q$ and higher $Q$ sides of Bragg peaks (006) and (00$\bar{6}$), and can be indexed to the (204) and (206) Bragg peaks, respectively.

In both crystals, the diffraction pattern is dominated by a majority phase with the \textit{I4/m} space group with Fe vacancies and a minority phase consisting of the high symmetry \textit{I4/mmm} space group with no vacancies at the Fe site and a weak $\sqrt{2}\times\sqrt{2}$ vacancy order at the K site. Shown in Fig. 2(a) are the (200) Bragg peak from the \textit{I4/mmm} minority phase and the (420) Bragg peak from the \textit{I4/m} majority phase in the \textit{hk0} plane. They are well-resolved given that the two phases have different lattice constants ($a/\sqrt{5}\sim$3.90 \AA \ in \textit{I4/m}, $a\sim$3.84 \AA \ in \textit{I4/mmm}), often difficult to see in powders. Shown in Figs. 2(c) are the powder integrated diffraction patterns obtained from the annealed and as-grown crystals in the vicinity of the ($\frac{1}{2}$ $\frac{1}{2}$ 0) superlattice peak. Even though this peak is observed in both diffraction patterns, it is significantly stronger and clearly above the background level in the as-grown crystal at $Q\sim1.16$ \AA $^{-1}$ but barely visible in the annealed sample. The ($\frac{1}{2}$ $\frac{1}{2}$ 0) peak is not as intense as the other superlattice features which suggests that the K-site vacancy is partially ordered in the \textit{I4/mmm} phase. The K-site vacancy order can break the symmetry of the centrosymmetric \textit{I4/mmm} to \textit{P4/mmm} or to an even lower symmetry depending on its out-of-plane stacking pattern. However, our out-of-plane diffraction data did not provide enough information to further confirm the symmetry. Single crystal refinement was performed on the \textit{hk0} plane data, and the results are summarized in Tables I and II, where space group \textit{P4/mmm} was used to refine the ($\frac{1}{2}$ $\frac{1}{2}$ 0) superlattice peak of the minority phase. The refinement yielded a volume fraction for the \textit{I4/mmm} phase of 18.4(3)\% in the annealed sample and about 31.6(3)\% in the as-grown. Given that there is less of the \textit{I4/mmm}, the presumed host of the SC state, in the annealed sample which is SC than in the as-grown sample which is NSC, it is questionable whether or not this is the phase in which superconductivity occurs. How the K vacancy order affects superconductivity is still a question. At the same time, the refinement indicates that the \textit{I4/m} phase is not fully ordered with the $\sqrt{5}\times\sqrt{5}\times1$ Fe vacancy ordered supercell. Shown in Fig 2(b) is a comparison of the integrated intensity of the (110)$_{I4/m}$ superlattice peak to the calculated intensity assuming a fully ordered Fe-vacancy sublattice with no occupancy at the Fe1 site. The experimental intensity is lower which suggests that even within the majority phase, two different Fe sublattices are present, a fully vacancy ordered one and a partially ordered (or disordered) one. The disordered Fe sublattice is described within the \textit{I4/mmm} symmetry, but it is indistinguishable from the ordered Fe sublattice because their lattice constants are unresolved in the experimental data.

\begin{figure}[ptb]
	\includegraphics[width=0.44\textwidth]{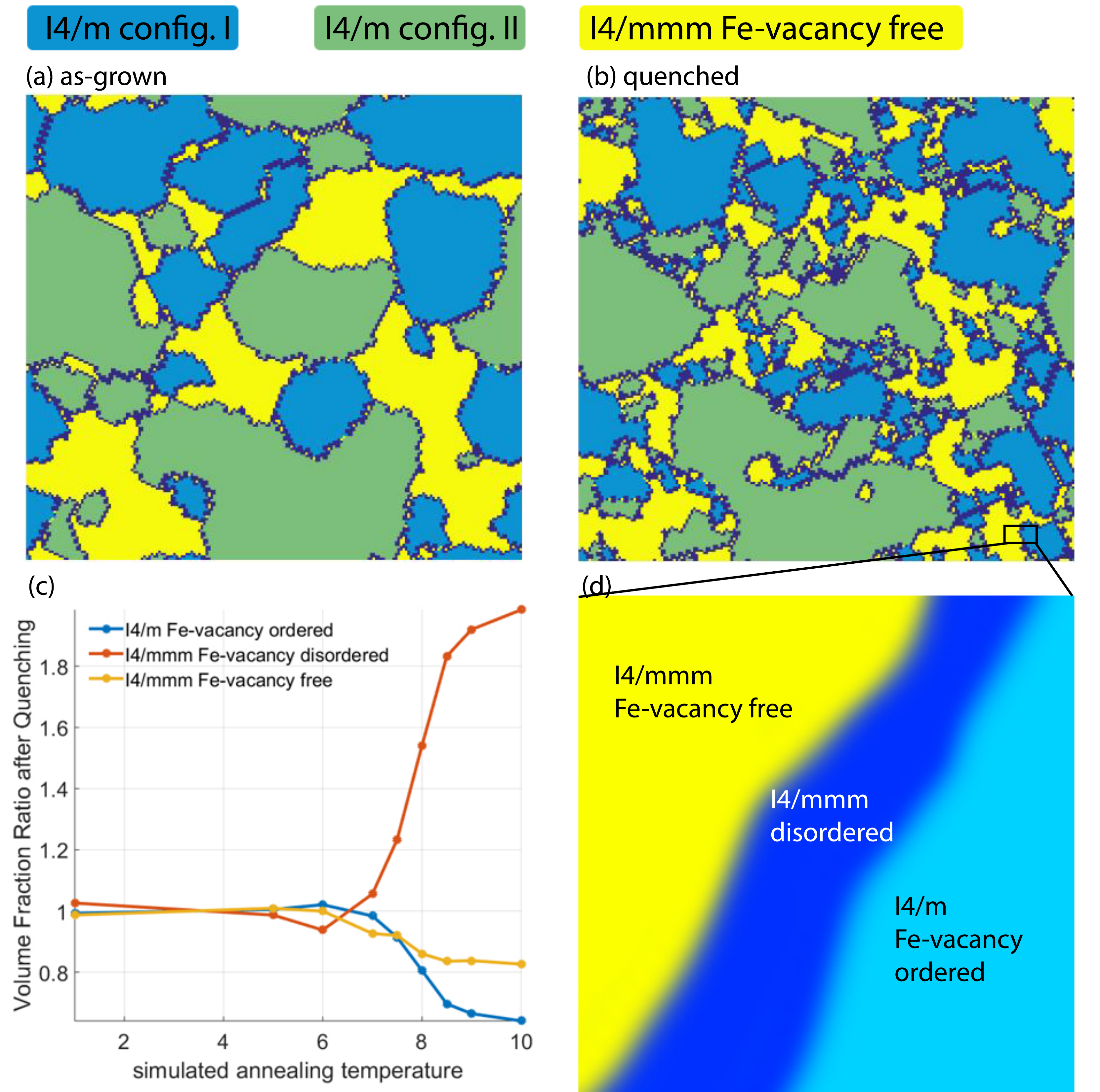} \caption{(a) and (b) The phase distribution of the simulated lattice before and after simulated annealing and quenching processes. (c) The volume fraction ratio of each phases before and after quenching. (d) A schematic plot demonstrating the three phases.}%
	\label{fig3}%
\end{figure}

To quantify the differences between the ordered and disordered phases, a Monte Carlo simulation on the Fe sublattice was performed. The Hamiltonian was designed to be Ising like with the following form: $H=\sum_{<21>}J_{21}\sigma_{i}\sigma_{j}+\sum_{<31>}J_{31}\sigma_{i}\sigma_{j}+\sum_{<11>}J_{11}\sigma_{i}\sigma_{j}+\sum_{<20>}J_{20}\sigma_{i}\sigma_{j}+\sum_{<22>}J_{22}\sigma_{i}\sigma_{j}$. Here the Ising variable $\sigma_{i}=1$ represents an Fe atom at site $i$ and $\sigma_{i}=-1$ stands for a vacancy at site $i$. The coupling constant between $\sigma_{i}$ and $\sigma_{j}$ were defined up to the 5th nearest neighbor. By setting $J_{11,20,22}>0$, $J_{21,31}<0$, the two $\sqrt{5}\times\sqrt{5}$ configurations will be energetically favored. For the Monte Carlo step, site swapping was employed instead of site flipping, in order to keep the vacancy ratio unchanged. When vacancies are less than 20 \%, regions with no Fe vacancies will form on the lattice, simulating the \textit{I4/mmm} phase. The as-grown sample was simulated by gradual cooling, then the system was heated to a high temperature, $T_{a}$, and cooled back down to simulate the annealed sample. The simulation results with 15 \% vacancies on a $300\times300$ lattice with $J_{11,20,22}=6$, $J_{21,31}=-1.5$ and $T_{a}=10$ are shown in Fig. 3. Before annealing, the two $\sqrt{5}\times\sqrt{5}$ configurations (blue and green) and the Fe vacancy free \textit{I4/mmm} phase (yellow) appear in big domains (Fig. 3(a)). After annealing, many small domains with the $\sqrt{5}\times\sqrt{5}$ structure form inside the previously vacancy free regions, breaking the \textit{I4/mmm} domains into smaller islands, creating more domain boundaries (Fig 3(b)). The simulated lattice is more homogeneous after the annealing process, in agreement with our SEM results\cite{junjie}. The volume ratio of the \textit{I4/mmm} phase also decreases after annealing, which agrees with our X-ray data.

\begin{table}[t]
\caption{Refined structure parameters for the \textit{I4/m} phase that includes both the Fe-vacancy disordered and Fe-vacancy ordered phases. Atomic position: K1, 2$a\ (0,0,0)$; K2, 8$h\ (x,y,0)$; Fe1, 4$d\ (0,\frac12,\frac14)$; Fe2, 16$i\ (x,y,0.2515)$; Se1, 4$e\ (\frac12,\frac12,0.1351)$; Se2, 16$i\ (x,y,0.1462)$. Out-of-plane coordinates are not refined, values are from ref\cite{wbao}. If not listed, the site occupancy (Occ.) is 1.}%
\label{tab:table1}
\begin{ruledtabular}
		\begin{tabular}{cccc}
		&		& SC 	& NSC\\
	\hline
		& $a$(\AA)	& 8.7261(7)	& 8.7243(5)\\
		& $c$(\AA)	& 14.108(4)	& 14.104(4)\\
	K1 	& Occ. 		& 0.716(8) 	& 0.79(2)\\
		& U$_{iso}$	& 0.016(2)	& 0.005(5)\\
	K2	& $x$		& 0.4009(4)	& 0.405(1)\\
		& $y$		& 0.1829(5)	& 0.173(1)\\
		& Occ.		& 0.716(8)	& 0.79(2)\\
		& U$_{iso}$	& 0.016(2)	& 0.005(5)\\
	Fe1	& Occ.		& 0.115(7)	& -0.03(2)\\
		& U$_{iso}$	& 0.0191(6)	& 0.019(2)\\
	Fe2	& $x$		& 0.1898(4)	& 0.1995(3)\\
		& $y$		& 0.0953(2)	& 0.0905(4)\\
		& Occ.		& 0.980(3)	& 1.00(1)\\
		& U$_{iso}$	& 0.0191(6)	& 0.019(2)\\
	Se1	
		& U$_{iso}$	& 0.0242(5)	& 0.015(1)\\
	Se2	& $x$		& 0.1166(2)	& 0.1117(4)\\
		& $y$		& 0.3021(2)	& 0.2952(3)\\
		& U$_{iso}$	& 0.0242(5)	& 0.015(1)\\
	\hline
	$\chi^2$	&			& 1.03		& 8.28\\
volume frac.&			& 81.6(3)\%	& 68.4(3)\%\\
		\end{tabular}
	\end{ruledtabular}
\end{table}

\begin{table}[t]
\caption{Refined structure parameters for the \textit{P4/mmm} Fe-vacancy free phase. Atomic position: K1, 1$a\ (0,0,0)$; K2, 2$e\ (\frac12,0,\frac12)$; K3, 1$c\ (\frac12,\frac12,0)$; Fe1, 8$r\ (x,x,0.25)$; Se1, 2$g\ (0,0,0.1456)$; Se2, 2$h\ (\frac12,\frac12,0.1456)$; Se3, 4$i\ (0,\frac12,0.3544)$. Out-of-plane coordinates are not refined, values are from ref\cite{wbao}. If not listed, the site occupancy (Occ.) is 1.}%
\label{tab:table2}
\begin{ruledtabular}
		\begin{tabular}{cccc}
			&		& SC 	& NSC\\
			\hline
			& $a$(\AA)	& 5.437(1)	& 5.433(1)\\
			& $c$(\AA)	& 14.230(7)	& 14.237(2)\\
		K1, K2 	
			& U$_{iso}$	& 0.031(4)	& 0.044(3)\\
		K3	& Occ.		& -0.03(4)	& 0.20(2)\\
			& U$_{iso}$	& 0.031(4)	& 0.044(3)\\
		Fe1	& $x$		& 0.2531(5) & 0.2505(3)\\
			& U$_{iso}$	& 0.0109(6)	& 0.0242(4)\\
		Se1, Se2, Se3	
			& U$_{iso}$	& 0.023(1)	& 0.0317(8)\\
		\hline
	$\chi^2$	&			& 1.26		& 0.62\\
volume frac.&			& 18.4(3)\%	& 31.6(3)\%\\
		\end{tabular}
	\end{ruledtabular}
\end{table}

The annealing process controls the phase distribution as seen in Fig 3 (c). With annealing, the Fe disordered \textit{I4/mmm} phase grows significantly over the \textit{I4/mmm} Fe vacancy-free and \textit{I4/m} Fe vacancy-ordered phases. Difficult as it is to separate the contribution of the \textit{I4/mmm} Fe disordered phase in the diffraction pattern, the difference between the experimental and calculated (110) superlattice peak intensities shown above in Fig 2(b) is indication that the \textit{I4/m} is not fully ordered, consistent with the calculation. How does this affect superconductivity? Our Monte Carlo simulation indicates that the annealing process increases the total area of the domain boundary where Fe vacancies tend to be randomized. The increase in the domain boundary walls is seen in Fig 3(b) while the length of the boundary walls increases as the domains get smaller. In a real sample this disorder can be enhanced by the local distortion at the domain boundaries due to different lattice constants of the two phases. It was previously shown using thin films of K$_{x}$Fe$_{2-y}$Se$_{2}$ \cite{li} that the superconducting phase appears when the \textit{I4/mmm} phase borders the \textit{I4/m} phase. The domain boundary forms a filamentary network of Fe vacancy disorder. Fe vacancy disorder suppresses the band structure reconstruction and raises the chemical potential without completely destroying the Fermi surface \cite{tom}. The Fe vacancy disorder can thus serve as effective doping and lead to superconductivity. This is consistent with our X-ray and simulation results and provides a connection to the transport properties of the two samples.

To conclude, the presence of competing degrees of freedom is a common theme in superconductors of interest today. \ In our system, the SC crystal is a multi-phase separated state just like the NSC crystal and what distinguishes the two is the extent of the Fe-vacancy disordered state. \ Our results offer contradictory evidence to the motion that the \textit{I4/mmm} phase with no Fe vacancies is the host of the superconductivity. On the contrary, the SC crystal tends to form more domain boundaries with the Fe-vacancy disordered phase sandwiched between the \textit{I4/mmm} vacancy free and the \textit{I4/m} vacancy ordered phases as seen in Fig 3(d), and very possibly leads to superconductivity in a filamentary form, in agreement with a reported SPEM study \cite{bianconi}. In this way we provide a reasonable understanding of the enhancement of superconductivity by annealing as well as the filamentary nature of the superconductivity in this compound, a common feature observed in other superconductors. \ The \textit{I4/m} and \textit{I4/mmm} domains are reminiscent of the charge paddles observed in cuprates \cite{bianconi2}.

The authors would like to acknowledge valuable discussions with W. Bao, T. Egami, A. Bianconi, and W. Ku. The work at the University of Virginia is supported by the U. S. Department of Energy, Office of Basic Energy Science, DE-FG02-01ER45927.

*Corresponding author

E-mail: louca@virginia.edu

\end{document}